# A bioinformatics pipeline for the identification of CHO cell differential gene expression from RNA-Seq data.

Craig Monger[1,2], Krishna Motheramgari[1,2], John McSharry[1], Niall Barron[2] and Colin Clarke[1*].

[1] National Institute for Bioprocessing Research and Training, Fosters Avenue, Blackrock, Co. Dublin, Ireland.
[2] National Institute for Cellular Biotechnology, Dublin City University, Dublin 9, Ireland.

**Running title:** An *in-silico* CHO cell RNA-Seq data analysis protocol.

**\*Corresponding Author:** Dr. Colin Clarke
*Tel. No.*: +353-1-2158100
*Fax. No.*: +353-1-2158116
*E-mail*: colin.clarke@nibrt.ie

**Abstract:** In recent years the publication of genome sequences for the Chinese hamster and Chinese hamster ovary (CHO) cell lines have facilitated study of these biopharmaceutical cell factories with unprecedented resolution. Our understanding of the CHO cell transcriptome, in particular, has rapidly advanced through the application of next-generation sequencing (NGS) technology to characterise RNA expression (RNA-Seq). In this chapter we present a computational pipeline for the analysis of CHO cell RNA-Seq data from the Illumina platform to identify differentially expressed genes. The example data and bioinformatics workflow required to run this analysis are freely available at www.cgcdb.org/rnaseq_analysis_protocol.html.

# 1 Introduction

Our understanding of Chinese hamster ovary (CHO) cell biology has dramatically improved in recent years bringing the promise of rational genetic engineering to enhance biopharmaceutical production closer to reality. The catalyst for these rapid advances has undoubtedly been the publication of genome sequences for multiple CHO cell lines and the Chinese hamster [1–3]. These data have had a broad impact on the field revealing, for instance, CHO cell line heterogeneity [2,4], improving proteomic characterisation [5], and enabling the use of genome editing technologies such as CRISPR-Cas9 [6]. The availability of genomic data has also improved the accuracy and decreased the cost of next generation sequencing based transcriptomics (RNA-Seq). The alignment of reads to a closely related species (*i.e.* mouse) or the deep RNA sequencing required to accurately reconstruct the transcriptome *de novo* is no longer necessary for CHO cell RNA-Seq.

In this chapter, we present an *in-silico* protocol for differential mRNA expression analysis from Illumina RNA-Seq data utilising the Chinese hamster genome as a reference sequence. The typical stages of a bioinformatics workflow are outlined (Figure 1) as well the commands required to perform each operation. To facilitate the reproduction of this analysis we have made both the example data and computer code freely available (www.cgcdb.org/rnaseq_analysis_protocol.html). The pipeline begins by illustrating the detection of common issues in raw sequencing data using *FASTQC* [7] and correcting those issues using *Trimmomatic* [8]. Pre-processed reads are aligned to the Chinese hamster reference genome (C_griseus_v1.0, RefSeq Assembly accession: GCF_000419365.1) with *HISAT2* [9] a fast splice-aware alignment algorithm (a pre-built *HISAT2* genome index is provided for this purpose). *RNASeqQC* [10] is used to determine the effectiveness of read mapping to the reference genome. Finally the number of reads aligning to annotated genes in the Chinese hamster genome is determined using *HTSeq* [11] and imported into the R statistical software environment where differential expression analysis is accomplished using the *DESeq2* [12] Bioconductor package.

# 2 Materials

## *2.1* Software installation

This bioinformatics pipeline is configured for the Linux operating system in order to ensure compatibility with widely used RNA-Seq data analysis software. Ubuntu 16.04 LTS (http://www.ubuntu.com) has been extensively tested for the analysis described in this chapter and is recommended for users unfamiliar with Linux due to its Windows-like desktop. The analysis workflow initially utilises standalone packages (Table 1) while the final stage is carried out within the R statistical software environment and the Bioconductor *DESeq2* package is used for differential expression analysis (*see* Note 1). A Bash script (*install_software.sh*) has been developed to automatically create the required directories, download and install each programme as well as any dependencies (*see* Note 2). This script ensures that the installation of each component of this protocol is straightforward and avoids compatibility issues that may occur in later stages of the analysis. Administrative privileges (i.e. *sudo*) are required to successfully run the analysis and the user will be prompted to provide a password when required. Users can download and execute the installation script by typing the following commands in the Linux terminal (*see* Note 3 & Note 4):

```
# script must be run from the user's home directory
cd $HOME
# Download the installation script
wget -N www.cgcdb.org/rnaseq_protocol/install_software.sh
# Execute the install script
bash install_software.sh
```

*2.2 Data download*

The example data utilised for this protocol originates from a previously published study focussed on the identification of receptors for TLQP-21, a peptide that affects energy metabolism and stress responses [13]. RNA-Seq was utilised to compare gene expression differences between CHO-K1 and CCL39 cells (derived from Chinese hamster lung tissue), which are responsive and non-responsive to TLQP-21 respectively. Total RNA sequencing on an Illumina HiSeq 2000 configured to acquire 76bp paired end reads was performed for 3 biological replicates of each cell line. These data are available for download on the NCBI Sequence Read Archive (accession: SRA096825). To enable the protocol described in this chapter to be carried out on a desktop computer the original data has been reduced (downsampled) to 2 million reads per sample (*see* Note 5). A Bash script (*download_data.sh*) is provided to download the RNA-Seq data for each sample along with additional resources required. The script automatically stores data in the required directories for analysis (Table 2).

```
# script must be run from the user's home directory
cd $HOME
# Obtain the data download script
wget -N www.cgcdb.org/rnaseq_protocol/download_data.sh
# Execute the install script
bash data_download.sh
```

## 3 Method

Once the required software and data have been downloaded a further two scripts are provided to automate each analysis stage in the pipeline. The first (*run_analysis.sh*) sequentially executes each program required for data pre-processing, read mapping to the Chinese hamster genome, alignment quality assessment and counts the number of reads aligning to features in the genome. The second script imports the read counts into the R software environment and performs differential expression analysis. Once complete, a PCA plot showing the global separation of samples based on their gene expression profiles as well as a file containing differentially expressed genes can be found in the "`$HOME/rnaseq_analysis/DESeq2_results`" directory. The main commands executed during the *run_analysis.sh* (Section 3.1-3.4) and *differential_expression.R* (Section 3.5) scripts are outlined below. The analysis scripts can be downloaded and executed as follows:

```
# script must be run from the user's home directory
cd $HOME
# Obtain the data download script
wget -N www.cgcdb.org/rnaseq_protocol/run_analysis.sh
# Execute pre-processing, reference genome alignment, alignment QC and counting
bash run_analysis.sh
# Obtain the differential expression R script
wget -N www.cgcdb.org/rnaseq_protocol/differential_expression_analysis.r
# Execute DESeq2 workflow
sudo Rscript differential_expression.r
```

**3.1 RNA-Seq raw data QC and pre-processing**

1. *Assess the quality of raw data using FastQC.* The command below analyses all (specified by the "*" character) FASTQ files in the raw directory folder and writes the result to the output folder specified by the "`--outdir`" flag (*see* Note 6).

```
$fastqc_directory/fastqc \
--outdir $HOME/rnaseq_analysis/FASTQC_output/raw_fastqc "$raw_data_directory"/*
```

2. *Pre-process the reads using Trimmomatic.* The command below carries out trimming based on the quality of each read in the sample (*see* Note 7) using a sliding window of 4 to assess the average Q score beginning at the 5' end of the read. If the average score falls below 20 the remainder of the read to the 3' end is removed. Following the trimming phase those reads with a minimum length < 25 nucleotides are also removed (*see* Note 8 & Note 9). *Trimmomatic* outputs pre-processed reads for the forward and reverse reads where read pairs are retained as well those where only one of the read pairs survived. The proportion of reads remaining in each sample is recorded in the *trimmomatic.log* file (*see* Note 10). Only paired reads are utilised for downstream stages of this pipeline.

```
java -jar $trimmotatic_directory/trimmomatic-0.36.jar PE \
"$raw_data_directory"/"$sampleName"_1.fq.gz \
"$raw_data_directory"/"$sampleName"_2.fq.gz \
"$preprocessed_data_directory"/paired/"$sampleName"_1.fq.gz \
"$preprocessed_data_directory"/unpaired/"$sampleName"_1.fq.gz \
"$preprocessed_data_directory"/paired/"$sampleName"_2.fq.gz \
"$preprocessed_data_directory"/unpaired/"$sampleName"_2.fq.gz \
SLIDINGWINDOW:4:20 MINLEN:25 >> "$preprocessed_data_directory"/trimmomatic.log
```

3. *Assess the quality of pre-processed data using FastQC.* Provides confirmation that issues have been corrected. Figure 2 illustrates the improvement in base quality scores at the 3' end of reads following *Trimmomatic* pre-processing.

```
$fastqc_directory/fastqc \
--outdir $HOME/rnaseq_analysis/FASTQC_output/raw_fastqc "$raw_data_directory"/*
```

### 3.2 Reference genome alignment

1. *Align sequence reads to the Chinese hamster genome.* The pre-processed data, where both read pairs have been retained, is aligned using *HISAT2* and the prebuilt C_griseus_v1.0 HISAT2 index. Alignments are outputted in the Sequence Alignment/Map (SAM) format. The "-x" option specifies the HISAT2 index, "-1" and "-2" are the input forward and reverse reads respectively. The "-S" option specifies the output file in SAM format (*see* Note 11).

```
hisat2 -x $hisat2_index/C_griseus_v1.0 --rg-id 1 --rg SM:Pool1 \
-1 $preprocessed_reads_directory/"$sampleName"_1.fq.gz \
-2 $preprocessed_reads_directory/"$sampleName"_2.fq.gz \
-S $mapped/"$sampleName".sam;
```

2. *Sort the SAM file and convert to BAM.* Each SAM format file produced during alignment is sorted based on location within each scaffold of the Chinese hamster genome and converted to its equivalent binary format BAM file. For the *Samtools view* command the "-bS" option specifies that the input is SAM format and that output should be BAM which is transferred to the *Samtools sort* command using the "|" character where the "-o" in the specifies the output file.

```
samtools view -bS $mapped/"$sampleName".sam | samtools sort - \
-o $mapped/"$sampleName".bam
```

3. *Remove the SAM file.* Once the BAM file is generated the SAM file created during *HISAT2* alignment is no longer required and deleted to reduce storage requirements.

```
rm $mapped/"$sampleName".sam
```

### 3.3 Reference genome alignment QC

1. *Create a FASTA index for the Chinese hamster genome.* A FASTA index file enables rapid access to sequence within Chinese hamster FASTA file.

```
samtools faidx $genome/GCF_000419365.1_C_griseus_v1.0_genomic.fasta
```

2. *Create a sequence dictionary for the Chinese hamster reference sequence.* "R=" specifies the FASTA sequence file from which to create dictionary while "O=" the output file. The resulting .dict file contains a list of names and sizes for each scaffold in the Chinese hamster genome.

```
java -jar $picard_directory/CreateSequenceDictionary.jar \
R=$genome/GCF_000419365.1_C_griseus_v1.0_genomic.fasta \
O=$genome/GCF_000419365.1_C_griseus_v1.0_genomic.dict
```

3. *Deduplicate pre-processed reads for RNASeqQC analysis.* The *Picard MarkDuplicates* program identifies duplicates reads in each dataset and retains only one of the duplicated reads for RNASeqQC. The "I" and "O" options specify input and output BAM files respectively while the "VALIDATION_STRINGENCY=SILENT" suppresses warning messages. A summary of the duplication process can be found in the .metric.txt file.

```
java -jar  $picard_directory/MarkDuplicates.jar \
I=$mapped/"$sampleName".bam \
O=$dedup_directory/"$sampleName".dup.bam \
M=$dedup_directory/"$sampleName".metric.txt VALIDATION_STRINGENCY=SILENT
```

4. *Calculate RNASeq metrics using RNASeqQC.* The *RNASeqQC* software calculates metrics including depth of coverage and GC bias. The '-s' option specifies a list of sample files and type to be analysed. A GTF annotation file is also supplied following the '-t' option that specifies the genomic locations of each mRNA (*see* Note 12). Table 3 illustrates selected *RNASeqQC* metrics utilised to evaluate the effectiveness of RNA-Seq.

```
java -jar $rnaseqqc_directory/RNA-SeQC_v1.1.8.jar \
-s $supplementary_files_directory/rnaseq_qc_sample_list.txt \
-t $supplementary_files_directory/ \
    GCF_000419365.1_C_griseus_v1.0_genomic.protein.coding.gtf \
-r $genome/GCF_000419365.1_C_griseus_v1.0_genomic.fasta \
-o $rnaseq_qc_output
```

### *3.4 Calculation of raw counts*

*1. Count the number of reads mapping to each gene in the Chinese hamster genome.* The *htseq-count* program is utilised to count the number of reads for each sample BAM file (the input file format is specified by the "-f" flag) mapping to each feature with a GFF format annotation file (*see* Note 13). Those features counted within the GFF file are specific by the "-i" flag in this case those with "gene" specified. The ">" character writes the output to a text file for further processing (*see* Note 14).

```
htseq-count –f bam –i gene –s no $mapped/"$sampleName".bam \
$genome/GCF_000419365.1_C_griseus_v1.0_genomic.protein.coding.gff \
> $count_directory/"$sampleName".counts
```

### *3.5 Differential gene expression analysis*

The remaining stages of this pipeline are executed within R and utilise the *DESeq2* Bioconductor package.

*1. Import the count data and create a DESeq2 object.* The count file location for each sample within the HTSeq_counts directory along with the cell type (e.g. CHO-K1 or CCL39) are placed in an R data frame. The data frame is utilised as input to the *DESeqDataSetFromHTSeqCount* function, which imports the count data and constructs a DESeq object. The cell type information is converted to a R factor variable for downstream sample comparisons.

```
# determine the names of the HTSeq count files
count_file_names <- grep("counts",list.files("HTSeq_counts"),value=T)

# set sampleConditions and sampleTable for experimental conditions
cell_type <- c("CCL39","CCL39","CCL39", "CHO-K1","CHO-K1","CHO-K1")

sample_information <- data.frame(sampleName = count_file_names,
                                 fileName   = count_file_names,
                                 condition  = cell_type)

# create a data deSeq2 object by reading the count files and assign cell type
DESeq_data <- DESeqDataSetFromHTSeqCount(sampleTable = sample_information,
                                         directory = "HTSeq_counts",
                                         design = ~condition)

# convert cell type to an R factor variable
colData(DESeq_data)$condition <- factor(colData(DESeq_data)$condition,
                                        levels = c('CCL39','CHO-K1'))
```

*2. Principal components analysis.* Principal components analysis (PCA) is a useful means to determine if the biological hypothesis underlying the experimental design is reflected in the global expression patterns observed from RNA-Seq analysis before progressing to differential expression analysis. The raw count data is first transformed using the regularised logarithm before *DESeq2*'s plotPCA function is used to carry out PCA, generate a scatter plot of the first vs the second principal component and label the points based on the cell type. The PCA plot for the RNA-Seq data utilised in this protocol illustrate clear separation of the CHO-K1 and CCL39 samples (Figure 3).

```
# transform the count data for PCA
rld <- rlogTransformation(DESeq_data, blind=T)

# create a PDF to save the figure
pdf("DESeq2_results/CHO-K1_v_CCL39.pdf")

# Plot Principal component 1 v Principal 2, label based on cell type
plotPCA(rld, intgroup="condition")

# save plot to PDF file
dev.off()
```

*3. Differential expression analysis using DESeq2.* Before conducting differential expression analysis a comparator sample is set using the relevel function, in this case the TLQP-21 non-responsive CCL39 cell line. The DESeq wrapper function is called on the DESeq object containing the raw counts, which executes the DESeq2 method which conducts size factor normalisation, removes genes with very low counts and calculates differential expression using built-in functions and adds the results to the existing object. The differential expression results are filtered based on fold change (≤ -2 or ≥ 2) as well as a Bonferroni adjusted p-value of < 0.05 (*see* Note 15) before ordering the results based on the degree of up or downregulation in CHO-K1 when compared to CCL39.

```
# set the comparator condition
DESeq_data$condition<-relevel(DESeq_data$condition, "CCL39")

# calculate differential expression using the DESeq wrapper function
DESeq_data <- DESeq(DESeq_data)

#set differential expression criteria
de_results<-results(DESeq_data, pAdjustMethod="bonferroni",
                    lfcThreshold=0, independentFiltering=T)

# order results by padj value (most significant to least)
sig_de_results <- subset(de_results, abs(log2FoldChange)> 1 & padj < 0.05)
sig_de_results <- sig_de_results[order(sig_de_results$log2FoldChange,
decreasing=T),]
```

*4. Annotate differentially expressed genes.* An annotation file is provided in the supplementary data that, when imported into the R environment using the read.csv function, can be used to add the gene name and GenBank ID to the differentially expressed genes identified by *DESeq2*.

```
# import annotation information
annotation_file<-
"supplementary_files/GCF_000419365.1_C_griseus_v1.0_genomic.protein.coding.csv"
annotation_info<-read.csv(annotation_file, row.names=1, header=T)

# identify annotation information for differentially expressed genes
sig_de_annotations <- annotation_info[rownames(sig_de_results),]

#combine annotation information and DESeq2 output
sig_de_results<-cbind(sig_de_annotations, as.data.frame(sig_de_results))
```

*5. Export differentially expressed genes to a CSV file.* Annotated significantly differentially expressed genes are written to a CSV file using the write.csv R function. Table 4 illustrates the final output of this pipeline for the 10 up and downregulated genes in CHO-K1 cells when compared with CCL39 cells.

```
write.csv(sig_de_results, row.names=T, file="DESeq2_results/CHO-K1_v_CCL39.csv",)
```

# 4 Notes

1. The pipeline in this chapter utilises selected software packages. Readers should note that there is a vast array of different bioinformatics software available. Alternatives to each stage of this pipeline are discussed in a recent review from our laboratory [14].

2. Bash scripts are provided for convenience and to ensure compatibility. Readers are encouraged to examine the contents of these files using a text-editor such as *gedit* or by printing the contents to the terminal using the following command: `more $HOME/install_software.sh`.

3. Readers should be aware that when copying and pasting commands directly from this article into the Linux terminal characters can be altered which will result in an error (e.g. single & double quotes as well as the "-" character).

4. To aid understanding of the computer code used in this protocol the following colour scheme is used: calls to software & R functions (blue), software options/flags/R function arguments (green). Variables (purple) are preceded by the "$" character and are used here as references or "shortcuts" to directories and sample filenames. Comments lines (grey) are preceded by the "#" character (ignored by Bash and R) and the "\" character specifies a new line within a Linux command (used for code readability).

5. Downsampling of the original data comprising of ~750 million paired-end reads to 2 million reads per sample is necessary to run this protocol on a standard desktop computer. Downsampling was carried out using *BBmap* [15]. The protocol has been tested on a 4GB computer with the downsampled data provided. ~50GB free hard drive space is required. Total runtime: ~10hrs (1 processor, 4GB RAM).

6. *FastQC* provides a user-friendly HTML based output that can be viewed in an Internet browser such as *Firefox*. The *run_analysis.sh* script will automatically open the browser and a new tab for each sample displaying the respective *FastQC* output.

7. The run_analysis.sh script uses a series of `while` loops to iterate over a list of sample names that are imported from a text file. Upon each iteration of the loop the $sample_name variable specifies the current file name in the list. The `while` loop completes when processing of each of the 6 samples has been carried out. In this chapter we focus on the commands within the loops, for those readers wishing to understand how a while loop is constructed in Bash should see the `run_analysis.sh` script (*see* Note 1).

8. The *Trimmomatic* quality threshold of 20 used here should be modified as appropriate, for instance, less stringent parameters can be used with data from newer sequencing insturments that yield higher quality reads.

9. The presence of adapter sequences is commonly encountered in raw RNA-Seq data. The example data utilised here did not have a significant degree of adapter contamination and therefore no adapter trimming was carried out. Readers should note, however, that Trimmomatic can also be used to remove adapter sequences as well as low quality bases. Adapter trimming can be specified using the ILLUMINACLIP option in conjunction with a FASTA file of the adapter sequences used for library preparation.

10. *gedit* can be used to view a summary of *Trimmomatic* or `more "$preprocessed_data_directory"/trimmomatic.log` can be used to print the log file to the terminal.

11. The read group parameters (--rg-id 1 --rg SM:Pool1) are not required by HISAT2. In this protocol arbitrary read groups are added to the SAM file to ensure compatibility with *Picard* and *RNA-SeQC* during later stages of the pipeline. These parameters can be removed during alignment if users do not wish to use *RNA-SeQC* during their analysis.

12. Conversion of the NICB GFF format annotation file was achieved using the *gffread* utility (http://ccb.jhu.edu/software/stringtie/gff.shtml). This file is supplied for convenience.

13. The GFF file utilised here contains 14,781 protein-coding regions selected from the original GCF_000419365.1_C_griseus_v1.0_genomic.gff file and is provided here for demonstration purposes.

14. The final 5 lines of each sample count file contain the overall metrics for the counting process including the number of reads that could not be assigned to a genome feature. The following command can be used to print the counting metrics: `tail -n 5 $count_directory/"$sampleName".counts`.

15. In this protocol we utilise the Bonferroni p-value adjustment for demonstration purposes, however in certain cases it can be an overly conservative measure of statistical significance. Readers should note that a number of alternative less stringent adjustment methods are available for the *DESeq2* package e.g. Benjamini-Hochberg.

**Acknowledgements**

The authors gratefully acknowledge funding from Science Foundation Ireland (grant refs: 13/SIRG/2084 and 13/IA/1963) and the eCHO systems Marie Curie ITN programme (grant ref: 642663).**References**
[1] Brinkrolf, K., Rupp, O., Laux, H., Kollin, F., et al., Chinese hamster genome sequenced from sorted chromosomes. *Nat Biotech* 2013, 31, 694–695.
[2] Lewis, N.E., Liu, X., Li, Y., Nagarajan, H., et al., Genomic landscapes of Chinese hamster ovary cell lines as revealed by the Cricetulus griseus draft genome. *Nat Biotech* 2013, 31, 759–765.
[3] Xu, X., Nagarajan, H., Lewis, N.E., Pan, S., et al., The genomic sequence of the Chinese hamster ovary (CHO)-K1 cell line. *Nat Biotech* 2011, 29, 735–741.
[4] Kaas, C.S., Kristensen, C., Betenbaugh, M.J., Andersen, M.R., Sequencing the CHO DCB11 genome reveals regional variations in genomic stability and haploidy. *BMC Genomics* 2015, 16, 160.
[5] Meleady, P., Hoffrogge, R., Henry, M., Rupp, O., et al., Utilization and evaluation of CHO-specific sequence databases for mass spectrometry based proteomics. *Biotechnol. Bioeng.* 2012, 109, 1386–1394.
[6] Ronda, C., Pedersen, L.E., Hansen, H.G., Kallehauge, T.B., et al., Accelerating genome editing in CHO cells using CRISPR Cas9 and CRISPy, a web-based target finding tool. *Biotechnol. Bioeng.* 2014, 111, 1604–1616.
[7] FASTQC, http://www.bioinformatics.babraham.ac.uk/projects/fastqc/.
[8] Bolger, A.M., Lohse, M., Usadel, B., Trimmomatic: a flexible trimmer for Illumina sequence data. *Bioinformatics* 2014, 30, 2114–2120.
[9] Kim, D., Langmead, B., Salzberg, S.L., HISAT: a fast spliced aligner with low memory requirements. *Nat. Methods* 2015, 12, 357–360.
[10] DeLuca, D.S., Levin, J.Z., Sivachenko, A., Fennell, T., et al., RNA-SeQC: RNA-seq metrics for quality control and process optimization. *Bioinformatics* 2012, 28, 1530–1532.
[11] Anders, S., Pyl, P.T., Huber, W., HTSeq--a Python framework to work with high-throughput sequencing data. *Bioinforma. Oxf. Engl.* 2015, 31, 166–169.
[12] Love, M.I., Huber, W., Anders, S., Moderated estimation of fold change and dispersion for RNA-seq data with DESeq2. *Genome Biol.* 2014, 15, 550.
[13] Hannedouche, S., Beck, V., Leighton-Davies, J., Beibel, M., et al., Identification of the C3a Receptor (C3AR1) as the Target of the VGF-derived Peptide TLQP-21 in Rodent Cells. *J. Biol. Chem.* 2013, 288, 27434–27443.
[14] Monger, C., Kelly, P.S., Gallagher, C., Clynes, M., et al., Towards next generation CHO cell biology: Bioinformatics methods for RNA-Seq-based expression profiling. *Biotechnol. J.* 2015, 10, 950–966.
[15] BBMap - Bushnell B. - sourceforge.net/projects/bbmap/
[16] Li, H., Handsaker, B., Wysoker, A., Fennell, T., et al., The Sequence Alignment/Map format and SAMtools. *Bioinformatics* 2009, 25, 2078–2079.
[17] Picard, http://broadinstitute.github.io/picard/.12. Conversion of the NICB GFF format annotation file was achieved using the *gffread* utility (http://ccb.jhu.edu/software/stringtie/gff.shtml). This file is supplied for convenience.

13. The GFF file utilised here contains 14,781 protein-coding regions selected from the original GCF_000419365.1_C_griseus_v1.0_genomic.gff file and is provided here for demonstration purposes.

14. The final 5 lines of each sample count file contain the overall metrics for the counting process including the number of reads that could not be assigned to a genome feature. The following command can be used to print the counting metrics: `tail -n 5 $count_directory/"$sampleName".counts`.

15. In this protocol we utilise the Bonferroni p-value adjustment for demonstration purposes, however in certain cases it can be an overly conservative measure of statistical significance. Readers should note that a number of alternative less stringent adjustment methods are available for the *DESeq2* package e.g. Benjamini-Hochberg.

**Acknowledgements**

The authors gratefully acknowledge funding from Science Foundation Ireland (grant refs: 13/SIRG/2084 and 13/IA/1963) and the eCHO systems Marie Curie ITN programme (grant ref: 642663).

**Figure legends**

**Figure 1: RNA-Seq bioinformatics protocol overview. (A)** Quality assessment of raw sequencing data and pre-processing of reads to correct issues including low base quality. **(B)** Alignment of reads to the Chinese hamster reference sequence and calculation of global mapping quality. **(C)** Counting reads aligning to individual genes. **(D)** Differential expression analysis.

**Figure 2: Base quality pre-processing. (A)** *FASTQC* boxplots of base qualities scores shows a significant portion of reads have quality values falling below 20 at all nucleotide positions in forward reads for the CHO-K1_1 sample. For each nucleotide the boxes and whiskers show where 25-75% and 10-90% of the quality scores lie respectively, with the red horizontal line indicating the median quality value. **(B)** Trimming and filtering using the *Trimmomatic* tool significantly improves base qualities across the reads.

**Figure 3: Principal component analysis.** PCA provides a global overview of transcriptomics data, identifying outlying samples and confirming that the biological hypothesis underlying the experimental design can be observed. In this example the CHO-K1 and CCL39 samples clearly separate following PCA enabling subsequent identification of differential expression.

**Table legends**

**Table 1: List of software.** Standalone software written in Java, Python and C is required for this analysis as well as the R statistical software environment and the *DESeq2* Bioconductor package. The *install_software.sh* script automates the process of installing each program while the *DESeq2* package is installed during the execution of the *differential_expression.R* script. The purpose of each program is provided with respect to the protocol described here (*see* Note 1).

**Table 2: RNA-Seq data and additional resources.** The *download_data.sh* script creates the required directories and automatically downloads data required for this protocol including the downsampled raw FASTQ format files and *HISAT2* index files for alignment as well as GFF annotation files specifying the location of features in the Chinese hamster genome. Supplementary files are also provided for the RNASeqQC and the annotation of *DESeq2* results.

**Table 3: Read alignment quality metrics.** The *RNA-SeQC* program outputs a selection of measures of alignment quality for each RNA-Seq sample. For the example data utilised in this protocol an average of 91% of the total reads were aligned to the Chinese hamster genome sequence (mapping rate). The number of mapped reads that spanned an exon-exon junction is shown (split reads). Over 80% of aligned reads assigned to genes (intragenic rate) and over 74% to exons (exonic rate) enabling the detection of 9,000 genes and 18,000 mRNAs in each sample. A average profiling efficiency (sequenced reads vs. exon mapped reads) of 68.3% was achieved.

**Table 4: DESeq2 differential expression output.** A total of 572 upregulated and 924 downregulated protein coding genes were identified. The 10 most up and downregulated genes are shown here. The corresponding symbol, name and GenBank ID are shown for each gene. DESeq2 outputs including the fold change observed in $\log_2$ scale as well as the p-value as well as the Bonferroni adjusted p-value.

# Tables

## Table 1

| Software | Purpose | URL | Citation |
|---|---|---|---|
| *FastQC* | Read quality control | http://www.bioinformatics.babraham.ac.uk/projects/fastqc/ | [7] |
| *Trimmomatic* | Read pre-processing | http://www.usadellab.org/cms/?page=trimmomatic | [8] |
| *HISAT2* | Read alignment to a reference sequence | https://ccb.jhu.edu/software/hisat2/manual.shtml | [9] |
| *Samtools* | Manipulation of SAM/BAM files | https://github.com/samtools | [16] |
| *RNA-SeQC* | Alignment quality assessment | http://www.broadinstitute.org/cancer/cga/rna-seqc | [10] |
| *Picard* | RNA-Seq read deduplication | http://broadinstitute.github.io/picard/ | [17] |
| *HTSeq* | Counting #reads aligned to genome feature | http://www-huber.embl.de/users/anders/HTSeq/ | [11] |
| *R* | Differential Expression | https://cran.r-project.org/ | NA |
| *DESeq2* | Differential Expression | https://bioconductor.org/packages/release/bioc/html/DESeq2.html | [12] |

## Table 2

| Filename | Contents |
|---|---|
| **rnaseq_raw_data.tar.gz** | ❖ Raw RNA-Seq data for forward and reverse reads. 3 biological replicates for the CHO-K1 and CCL39 samples. |
| **hisat_index.tar.gz** | ❖ Pre-built HISAT2 index files for the Chinese hamster genome. |
| **C_griseus_v1.0.genomic.tar.gz** | ❖ Chinese hamster genome FASTA file.<br>❖ Chinese hamster GFF annotation file for protein coding genes. |
| **Supplementary_data.tar.gz** | ❖ RNASeqQC sample information.<br>❖ Chinese hamster GTF annotation file for protein coding genes.<br>❖ CSV file containing for annotation of differentially expressed genelist. |

Table 3

| Sample | Mapped | Split Reads | Mapping Rate (%) | Genes Detected | Transcripts Detected | Intragenic Rate (%) | Exonic Rate (%) | Intronic Rate (%) | Intergenic Rate (%) | Expression Profiling Efficiency (%) |
|---|---|---|---|---|---|---|---|---|---|---|
| **CCL39_1** | 2,263,342 | 491,468 | 89.0 | 9,342 | 18,138 | 80.8 | 74.4 | 6.4 | 19.2 | 66.2 |
| **CCL39_2** | 2,355,486 | 534,102 | 93.1 | 9,340 | 18,088 | 82.3 | 77.1 | 5.2 | 17.7 | 71.7 |
| **CCL39_3** | 2,267,567 | 494,761 | 89.7 | 9,376 | 18,215 | 80.8 | 74.3 | 6.4 | 19.2 | 66.7 |
| **CHO-K1_1** | 2,283,769 | 512,589 | 90.4 | 9,251 | 18,125 | 81.4 | 75.7 | 5.8 | 18.5 | 68.4 |
| **CHO-K1_2** | 2,314,711 | 505,472 | 91.7 | 9,292 | 18,173 | 81.1 | 74.3 | 6.9 | 18.9 | 68.1 |
| **CHO-K1_3** | 2,428,589 | 535,162 | 92.2 | 9,293 | 18,163 | 81.3 | 74.5 | 6.9 | 18.6 | 68.7 |

Table 4

| Gene Symbol | Gene Name | GenBank ID | Base Mean | $\log_2$ Fold Change | P-value | Adjusted P-value |
|---|---|---|---|---|---|---|
| **Slc45a1** | solute carrier family 45 member 1 | 100750542 | 277.79 | 8.28 | $6.79 \times 10^{-48}$ | $7.13 \times 10^{-44}$ |
| **Ccnd2** | cyclin D2 | 100771544 | 431.30 | 8.22 | $9.86 \times 10^{-66}$ | $1.04 \times 10^{-61}$ |
| **Eng** | endoglin | 100757433 | 641.12 | 8.10 | $6.57 \times 10^{-92}$ | $6.91 \times 10^{-88}$ |
| **Mmp9** | matrix metallopeptidase 9 | 100770707 | 166.28 | 7.38 | $3.30 \times 10^{-40}$ | $3.47 \times 10^{-36}$ |
| **Trim44** | tripartite motif containing 44 | 100757190 | 175.97 | 7.23 | $5.61 \times 10^{-43}$ | $5.9 \times 10^{-39}$ |
| **Dcaf12l2** | DDB1 and CUL4 associated factor 12-like 2 | 100764386 | 99.35 | 6.71 | $1.42 \times 10^{-31}$ | $1.49 \times 10^{-27}$ |
| **Fam134b** | family with sequence similarity 134 member B | 100766605 | 73.70 | 6.61 | $1.24 \times 10^{-26}$ | $1.30 \times 10^{-22}$ |
| **Ppp1r36** | protein phosphatase 1 regulatory subunit 36 | 100765040 | 108.81 | 6.40 | $3.93 \times 10^{-35}$ | $4.13 \times 10^{-31}$ |
| **Tbc1d16** | TBC1 domain family member 16 | 100754815 | 56.70 | 6.28 | $2.28 \times 10^{-23}$ | $2.39 \times 10^{-19}$ |
| **Hoxd8** | homeobox D8 | 100758537 | 70.42 | 6.26 | $2.45 \times 10^{-26}$ | $2.58 \times 10^{-22}$ |
| **Thbs2** | thrombospondin 2 | 100752022 | 261.34 | -7.89 | $1.44 \times 10^{-52}$ | $1.51 \times 10^{-48}$ |
| **Slc25a4** | solute carrier family 25 member 4 | 100751779 | 222.01 | -7.91 | $4.51 \times 10^{-48}$ | $4.74 \times 10^{-44}$ |
| **Cdh2** | cadherin 2 | 100689204 | 185.05 | -7.93 | $1.70 \times 10^{-42}$ | $1.78 \times 10^{-38}$ |
| **Cdh11** | cadherin 11 | 100767551 | 196.15 | -8.00 | $1.11 \times 10^{-43}$ | $1.17 \times 10^{-39}$ |
| **Rspo3** | R-spondin 3 | 100755164 | 250.24 | -8.21 | $1.12 \times 10^{-45}$ | $1.18 \times 10^{-41}$ |
| **Grb10** | growth factor receptor bound protein 10 | 100760784 | 532.50 | -8.45 | $7.24 \times 10^{-73}$ | $7.61 \times 10^{-69}$ |
| **Col1a2** | collagen type I alpha 2 | 100766269 | 617.51 | -9.04 | $1.35 \times 10^{-75}$ | $1.42 \times 10^{-71}$ |
| **Ftl** | ferritin- light polypeptide | 100753846 | 1049.17 | -9.07 | $3.44 \times 10^{-59}$ | $3.61 \times 10^{-55}$ |
| **Fabp4** | fatty acid binding protein 4 | 100760812 | 584.03 | -9.16 | $1.54 \times 10^{-70}$ | $1.62 \times 10^{-66}$ |
| **Bgn** | biglycan | 100771022 | 901.38 | -9.20 | $4.42 \times 10^{-93}$ | $4.64 \times 10^{-89}$ |

## Figures

## Figure 1

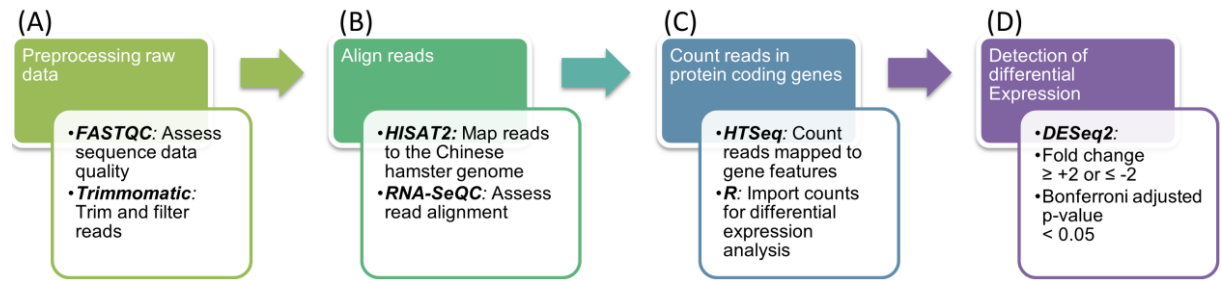

**Figure 2**

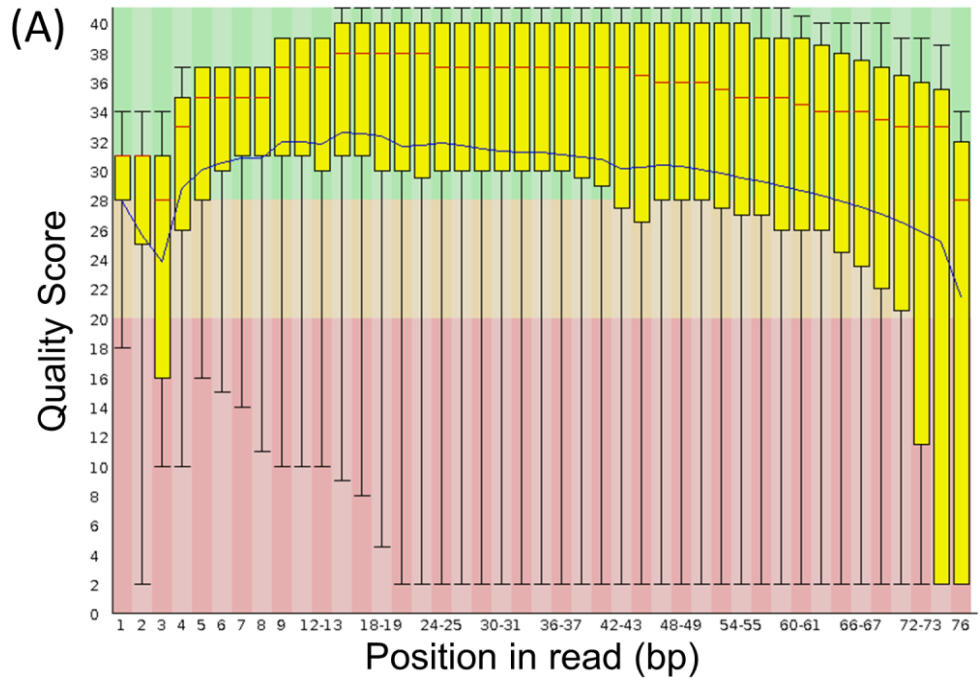

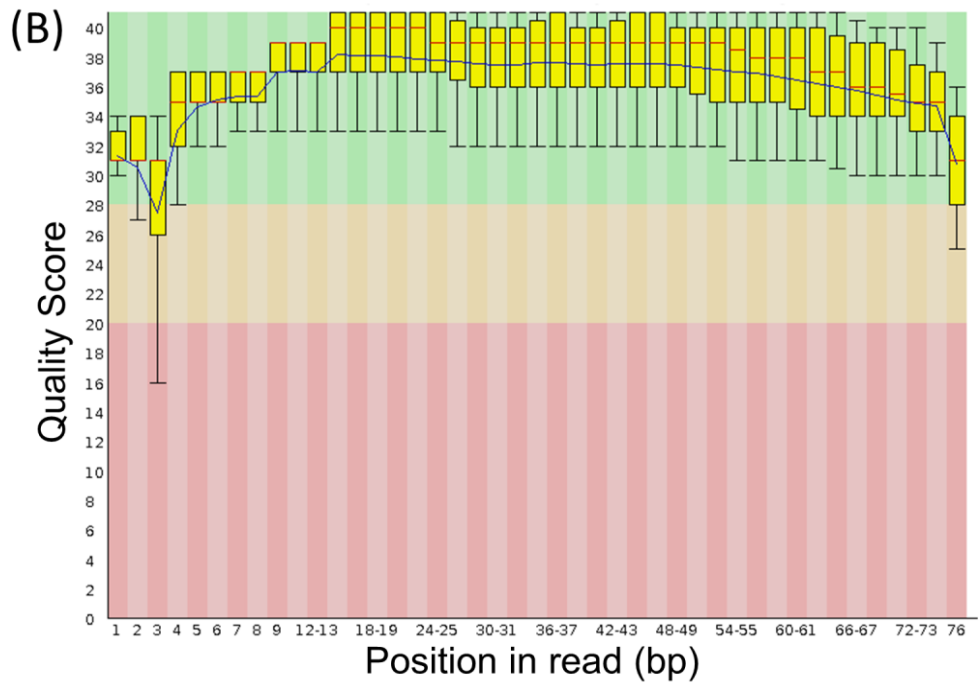

**Figure 3**

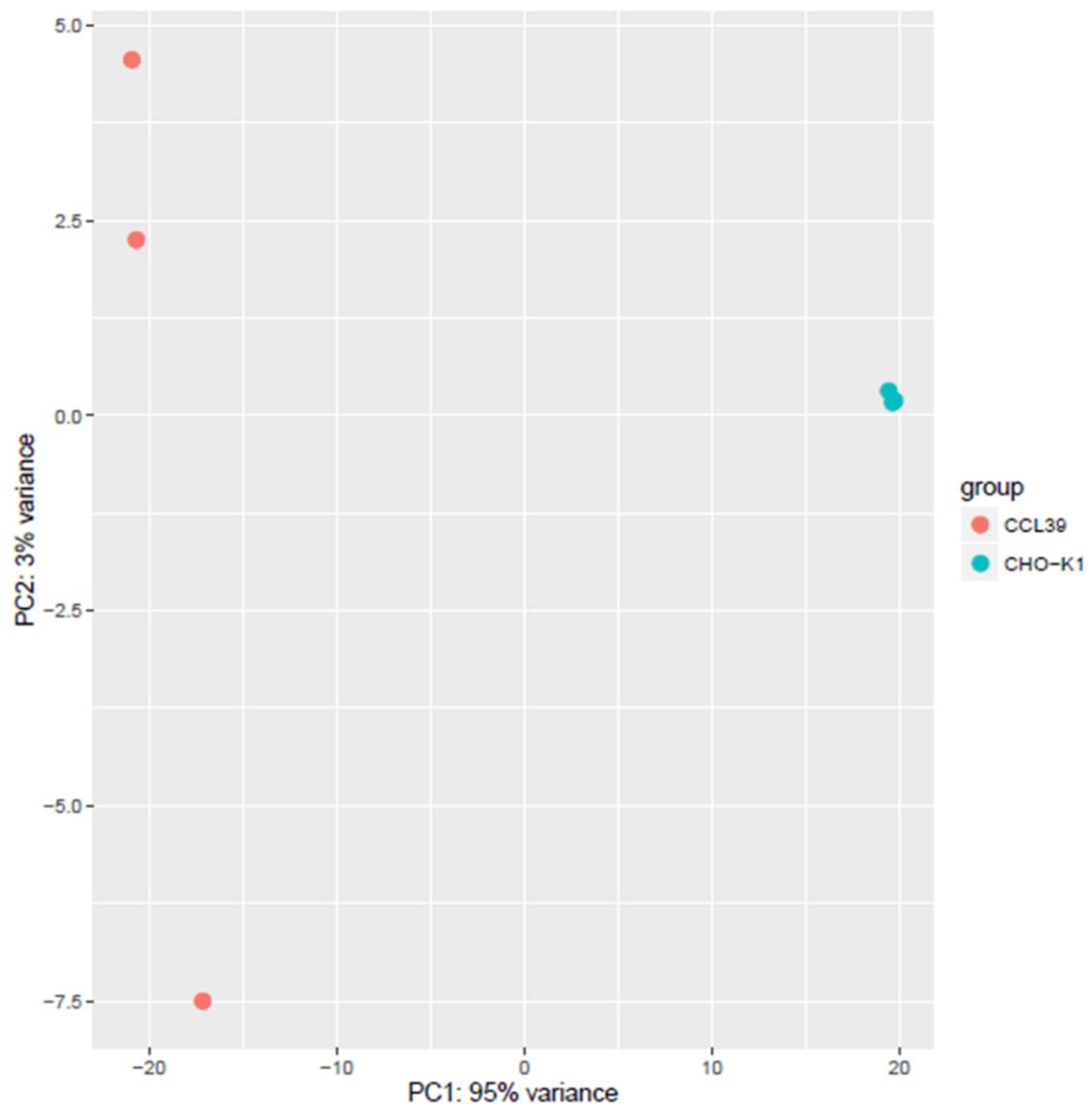